\documentclass[twocolumn]{aastex62}

\received{February 21}
\revised{July 21}
\accepted{July 29, 2019}
\submitjournal{ApJ}

\shorttitle{MIR-variable MYSOs}
\shortauthors{Uchiyama \& Ichikawa}

\usepackage{amsmath}
\usepackage{color}
\usepackage{url}
\usepackage{ulem}
\usepackage{multirow}
\usepackage{longtable}
\usepackage{natbib}
\usepackage{graphicx}
\usepackage{epstopdf}

\begin{document}

\title{\textit{WISE} Discovery of Mid-infrared Variability in Massive Young Stellar Objects}

\correspondingauthor{Mizuho Uchiyama}
\email{mizuho.uchiyama@nao.ac.jp}

\author[0000-0002-6681-7318]{Mizuho Uchiyama}
\affil{National Astronomical Observatory of Japan, 2-21-1 Osawa, Mitaka, Tokyo 181-8588, Japan}

\author[0000-0002-4377-903X]{Kohei Ichikawa}
\affil{Frontier Research Institute for Interdisciplinary Sciences, Tohoku University, Sendai 980-8578, Japan}
\affil{Astronomical Institute, Graduate School of Science
Tohoku University, 6-3 Aramaki, Aoba-ku, Sendai 980-8578, Japan}

\begin{abstract}

We systematically investigate the mid-infrared (MIR; $\lambda>3 ~\mu$m) time variability of uniformly selected $\sim800$ massive young stellar objects (MYSOs) from the Red MSX Source (RMS) survey. Out of the 806 sources, we obtain reliable 9-year-long MIR magnitude variability data of 331 sources at the 3.4~$\mu$m (W1) and 4.6~$\mu$m (W2) bands by cross-matching the MYSO positions with \textit{ALLWISE} and \textit{NEOWISE} catalogs.
After applying the variability selections using \textit{ALLWISE} data, we identify 5 MIR-variable candidates. 
The light curves show various classes, with the periodic, plateau-like, and dipper features. 
Out of the obtained two color-magnitude diagram of
W1 and W1$-$W2, one shows
``bluer when brighter and redder when fainter'' trends in variability, suggesting change in extinction or accretion rate.
Finally, our results show that G335.9960$-$00.8532 (hereafter, G335) has a periodic light curve, with a  $\approx 690$-day cycle. Spectral energy density model fitting results indicate that G335 is a relatively evolved MYSO; thus, we may be witnessing the very early stages of a hyper- or ultra-compact \ion{H}{2} region, a key source for understanding MYSO evolution.
 
\end{abstract}


\section{Introduction}

Massive stars have a significant role in star formation activities and metal enrichment in galaxies and, hence, the evolution of galaxies in the universe \citep[e.g.,][]{Tan14,ZY07}.
The powerful outflows of massive stars during the star-forming phase and subsequent expansion of \ion{H}{2} regions trigger next-generation star formation.
At the end of their evolution, these stars supply heavy elements to circumstellar regions through stellar winds and supernova explosions,
resulting in enriched chemical environments \citep[e.g.,][]{noz07}.
However, the detailed mass growth processes have yet to be resolved
due to observational difficulties.
The birthplace of such massive stars, called massive young stellar objects (MYSOs), are deeply obscured by dense gas and dust, preventing observation in the optical and even
sometimes near-infrared (NIR) regimes \citep{GK09}.
In addition, because they are relatively rare
and located at far distances ($>1$~kpc), 
it is difficult to spatially resolve the physical and geometrical structures within $<10$~AU \citep{Tan14}, where
mass growth phenomena occur.
Notably, these features are smaller than the currently achievable highest spatial resolution (20 mas at 230 GHz in the Atacama Large Millimeter/submillimeter Array (ALMA) extended configuration).
Very long baseline interferometry (VLBI) observations, such as \citet{Motogi16} in the H$_2$O 22.2~GHz band, have to date been the only means to resolve such compact structures, in which they mostly non-thermal emissions, such as masers, have been detected.
Physical parameters and structures of MYSOs, such as disk gas/dust density and temperature structure, stellar effective temperature and radius, and mass accretion processes, are still largely unknown. Thus, alternative observational methods are required to determine the properties of the spatially compact inner structures of MYSOs.

Variability studies may provide a way to address such difficulties. 
Observations in both the optical and NIR bands of low-mass young stellar objects (YSOs) have been conducted over the last 50 years via the small extinction of dust with $A_V<10$ \citep{MB99}. Such optical and NIR variability studies involving the emission lines and continuum of lower-mass YSOs have proven to be powerful tools for deciphering the physics of star formation and pre-main-sequence stellar evolution, such as the existence of rotating cool/hot spots \citep{BB89}, infalling/rotating dusty objects \citep{Cody14}, and accretion rate variation \citep{Audard14}. 
Recent intensive, large and deep mid-infrared (MIR; $\lambda>3 ~\mu$m) variability surveys of low-mass YSOs have enabled statistical studies, such as the Young Stellar Object VARiability (YSOVAR) project with \textit{Spitzer} \citep{Morales11}.
This statistical study revealed that flux variability in low-mass YSOs is ubiquitous in the optical, NIR, and MIR regimes.
They also categorized the causes of flux variability, such as stellar activity (e.g., cool spots), the density structure of disk, and mass accretion processes (e.g., hot spots, dust infalling, and episodic accretion), based on time and color variability characteristics \citep{Gunther14}.

Despite their importance, MYSOs have not been investigated extensively, especially in 
terms of IR thermal emission.
However, recently, 
high-amplitude ({$\Delta K_{\rm s} > 1 \ $mag}) year-scale variability in the NIR $K_{\rm s}$ was
reported for the first time, in 13 MYSOs using 
the Vista Variables in Via Lactea (VVV) survey data \citep{Kumar16}.
Following that study, lower-amplitude variables, approaching $\Delta K_{s} > 0.15 \ $mag,
have been detected in 190 out of 718 MYSO candidates in VVV survey data \citep{Teixeira18}.
These studies suggest that the variability is also common for MYSOs.
Recent theoretical studies also indicated that variable accretion rate and luminosity are common during mass growth phase of MYSOs 
\citep[e.g.,][]{Meyer19} and support the above observational results.
However, sample sets are still limited in these works, due to the  survey area of the Galactic center, 520 degree$^2$, and the difficulty in detection, as most MYSOs are still highly obscured in the NIR \citep{GK09}.
Thus, a flux variability survey that 1) covers a much more extensive area of the sky 
and 2) includes wavelengths longer than those of the NIR band
is essential to the study of the properties of massive star formation and mass growth processes.

In this paper, we report the first discovery of MIR-variable MYSOs using the the RMS survey MYSO catalog,
by cross-matching the MIR counterparts with the
multi-epoch \textit{ALLWISE} archives.
We also discuss the physical properties of detected-variable MYSOs using \textit{ALLWISE} and \textit{NEOWISE} data and their possible origins, based on the obtained light curves and color-magnitude diagrams (CMDs).

\begin{figure}
\begin{center}
\includegraphics[width=0.95\linewidth]{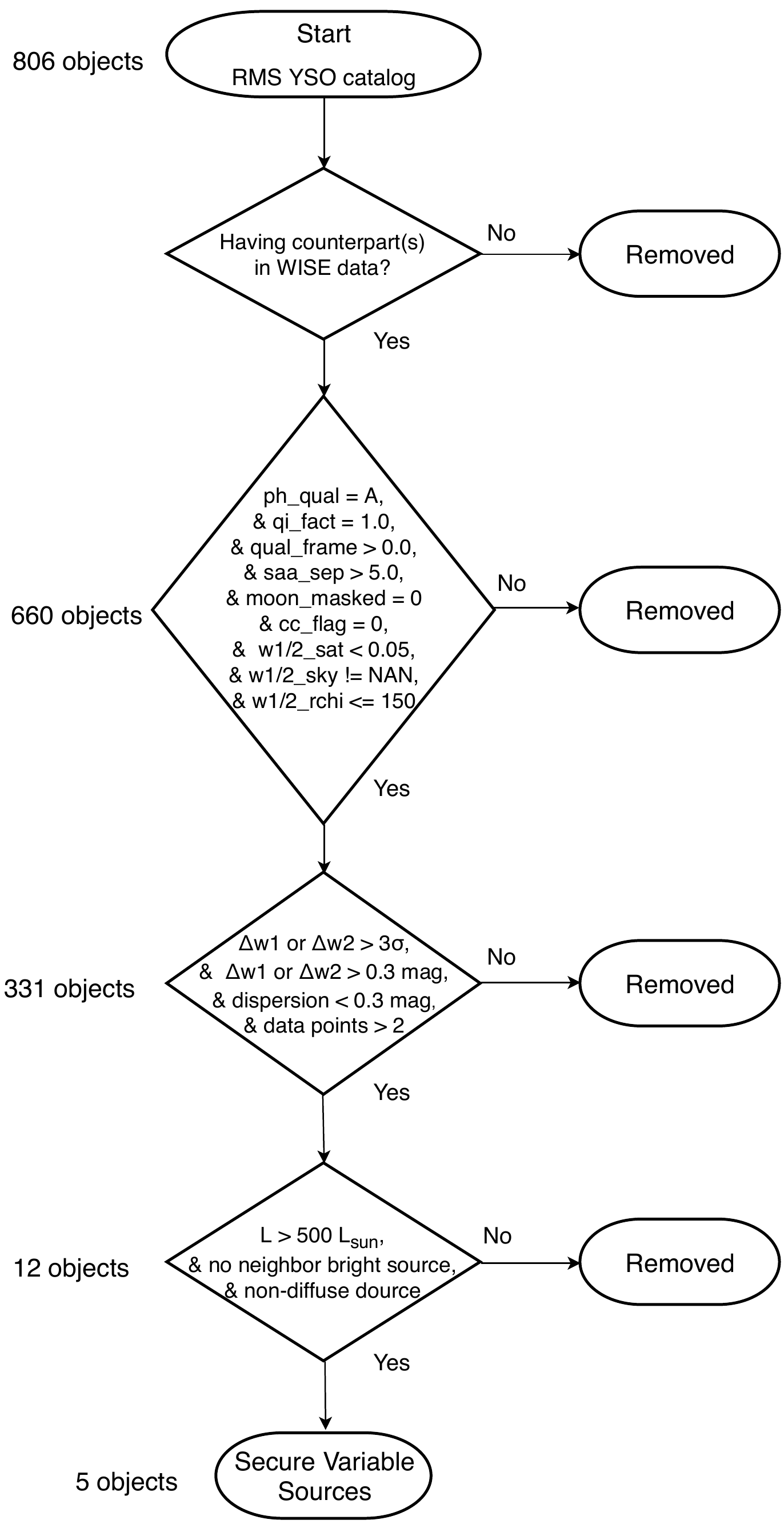}~
\caption{Summary flow chart of the sample selection process. The detail of each process is 
discussed in Section~\ref{sec:sample}.
}
\label{fig:flowchart}
\end{center}
\end{figure}

\section{Sample}\label{sec:sample}

Here we describe the sample in this study and the selection
process to find MIR-variable sources.
The flow chart of sample selection is compiled in Figure~\ref{fig:flowchart}.

\subsection{RMS Database}
Our initial sample was from
the RMS MYSO Database \citep{Lumsden13},
which is a multi-wavelength Galactic-plane MYSO survey
based on MSX IR satellite data archived
at 8.28, 12.13, 14.65, and 21.3~\micron~\citep{Egan03}.
It is the largest statistical catalog of young massive protostars with MIR detection, and the most suitable one for our purpose, i.e., searching for MIR-variable MYSOs in large samples.
It covers almost the entire Galactic plane, with the exception of the extremely crowded region of $10 \deg < l < 350 \deg$, with a latitude of $|b|<5 \deg$ \citep{Egan03}.
In this catalog, 806 MYSOs are listed by multi-color selection using the MSX and 2MASS data archives with criteria of $F_{21} > 2 F_{8}$, $F_{21} > F_{14}$, $F_{14} > F_{8}$, $F_{8} > 5 F_{K}$, and $F_{K} > 2 F_{J}$ \citep{Lumsden13,Lumsden02}.
These criteria effectively remove contaminants,
such as evolved stars with dust shells.
The RMS survey team also conducted follow-up radio observations;
any sources detected at 5~GHz above 10~mJy
were considered to be an \ion{H}{2} region or planetary nebulae \citep{Urquhart07_hii_s,Urquhart09_hii_n,Lumsden13};
hence, the contamination from the \ion{H}{2} region/planetary nebulae should be minimal in the data.
There are some known MYSOs associated with jets and detection at radio frequencies \citep[e.g.,][]{guz10,Lumsden13}.
However, as all are well below 10~mJy, the catalog accounts for such MYSOs
based on the radio detection criteria specified above.

\subsection{Cross-matching of the RMS Catalog with \textit{WISE}} \label{sec:matching}

We determined the MIR (band W1--W4; 3.4--22~$\mu$m) counterparts of the RMS MYSOs through positional
matching with \textit{ALLWISE} \citep[for all four bands;][]{wri10} and \textit{NEOWISE} \citep[only for W1 and W2; ][]{mai11,mai14}. 
We used NEOWISE 2019 Data Release version\footnote{More detailes information, see \url{http://wise2.ipac.caltech.edu/docs/release/neowise/neowise_2019_release_intro.html}} for the following analysis.
Hereafter, we will refer to both IR datasets as \textit{WISE}.

In this study, we applied a cross-matching radius
of 2 arcsec, based on the positional accuracy obtained with the 2MASS catalog \citep[see also][]{ich12, ich17a}.
When multiple \textit{WISE} objects satisfied this criterion, we removed them from our sample due to unreliable PSF photometry in WISE data reduction pipelines.
After this matching, 660 MYSOs showed corresponding \textit{WISE} objects.
We only used sources with flux quality \texttt{ph\_qual=A},
with a signal-to-noise ratio larger than 10.0,
and applied the single-exposure images of \texttt{qi\_fact=1.0},
which have the best image quality.
To select data which was not affected from artificial source contamination, South Atlantic Anomaly, or moon light, \texttt{qual\_frame>0.0}, \texttt{saa\_sep>5.0}, and \texttt{moon\_masked=0} were applied.
We also checked for sources of contamination and/or biased flux
in the vicinity of an image artifact (e.g., diffraction spikes, scattered-light halos,
and/or optical ghosts), using the contamination flag \texttt{cc\_flags}.
We removed sources containing a contamination flag with any letters, but retained sources flagged as \texttt{cc\_flags=0}
that are known to be unaffected by the known artifacts.
We then selected the sources with saturated pixels less than 5\% within the photometric apertures, flagged as \texttt{w1/2\_sat<0.05} to remove unreliable photometry with heavy saturation. 
Saturation begins to occur for point sources brighter than $\sim$ 8 and 7 mags in W1 and W2 bands, respectively \footnote{the details are compiled in
\url{http://wise2.ipac.caltech.edu/docs/release/allsky/expsup/sec2_4ci.html}}
We removed sources which failed background sky fitting or PSF profile fitting was not good, flagged as \texttt{w1/2\_sky=NAN} or \texttt{w1/2\_rchi2>150}.
These selection process finally leaves 331 sources, as shown in
Figure~\ref{fig:flowchart}.

\subsection{Variability Selection} \label{sec:variability}

MIR variability signals were checked for the selected 331 sources having reliable multi-epoch \textit{WISE} photometries.
In this study,
 the only \textit{ALLWISE} data were employed for MIR variability selections, due to the stability and accuracy of the photometries. In addition, we utilized only W1- and W2- band data, as multi-epoch data for these bands are readily available. Through this process,
objects with clear MIR variability signals were chosen, and hence note that sources with possible variability might have been missed in our study.

\begin{figure*}
\begin{center}
\includegraphics[width=0.33\linewidth]{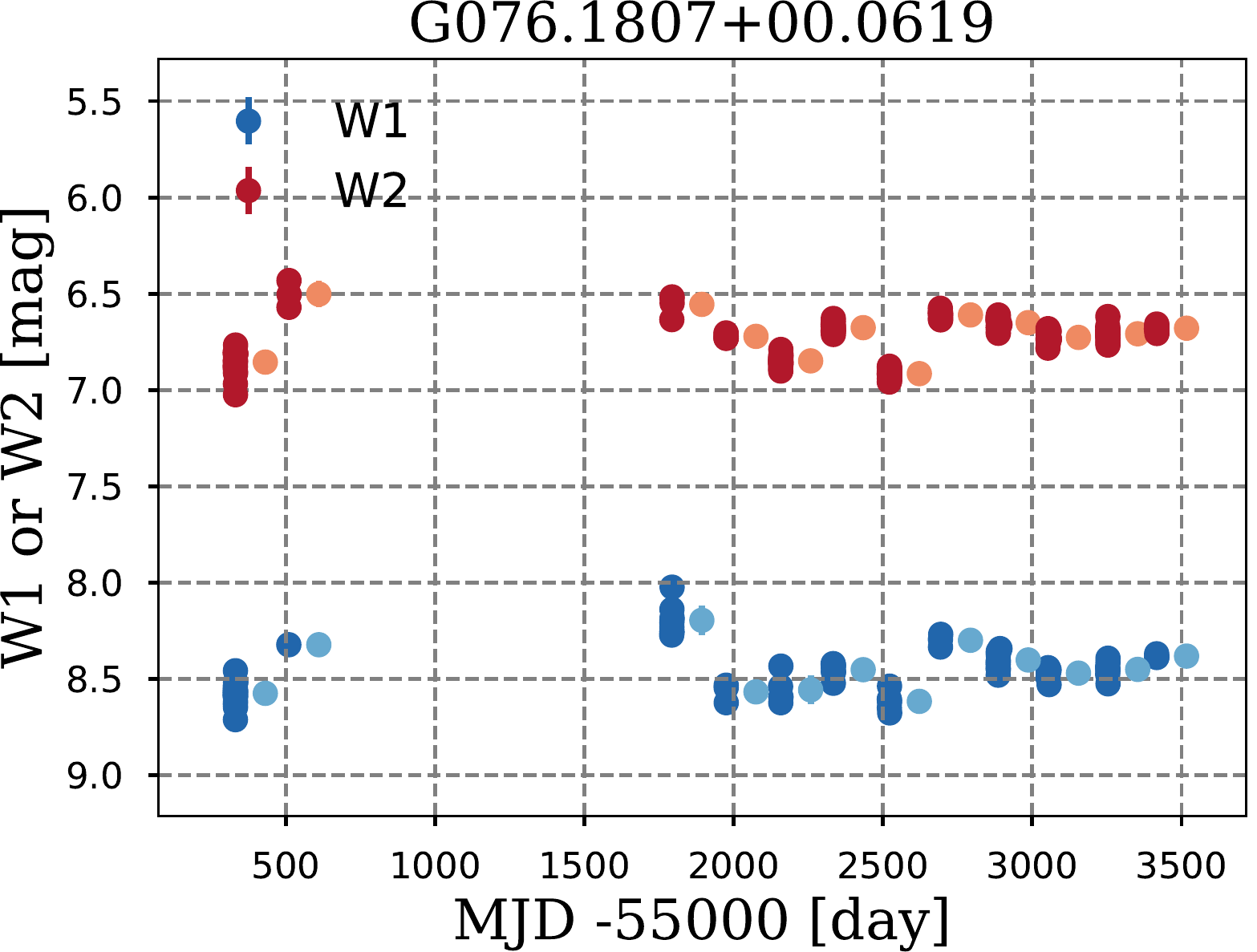}~
\includegraphics[width=0.33\linewidth]{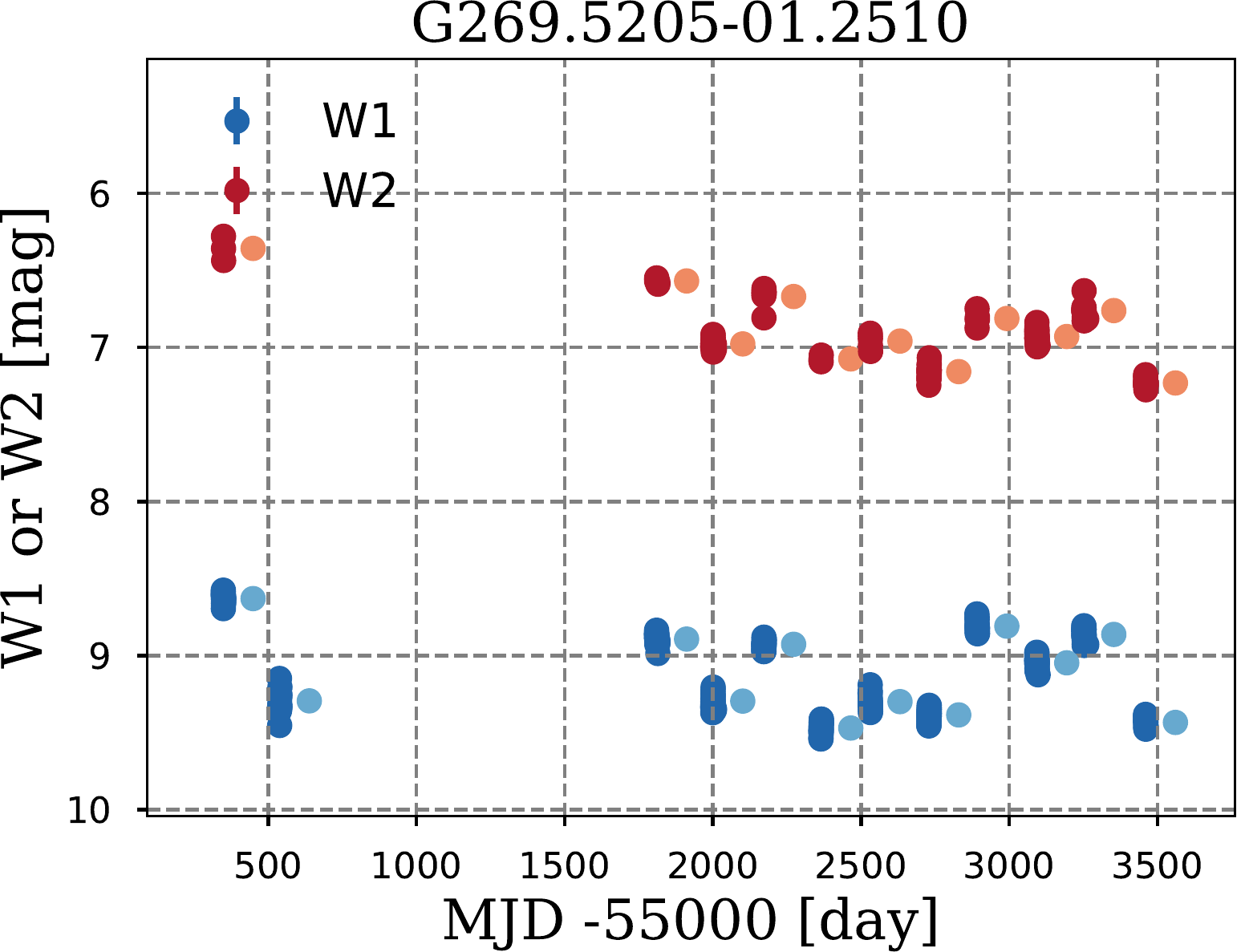}~
\includegraphics[width=0.33\linewidth]{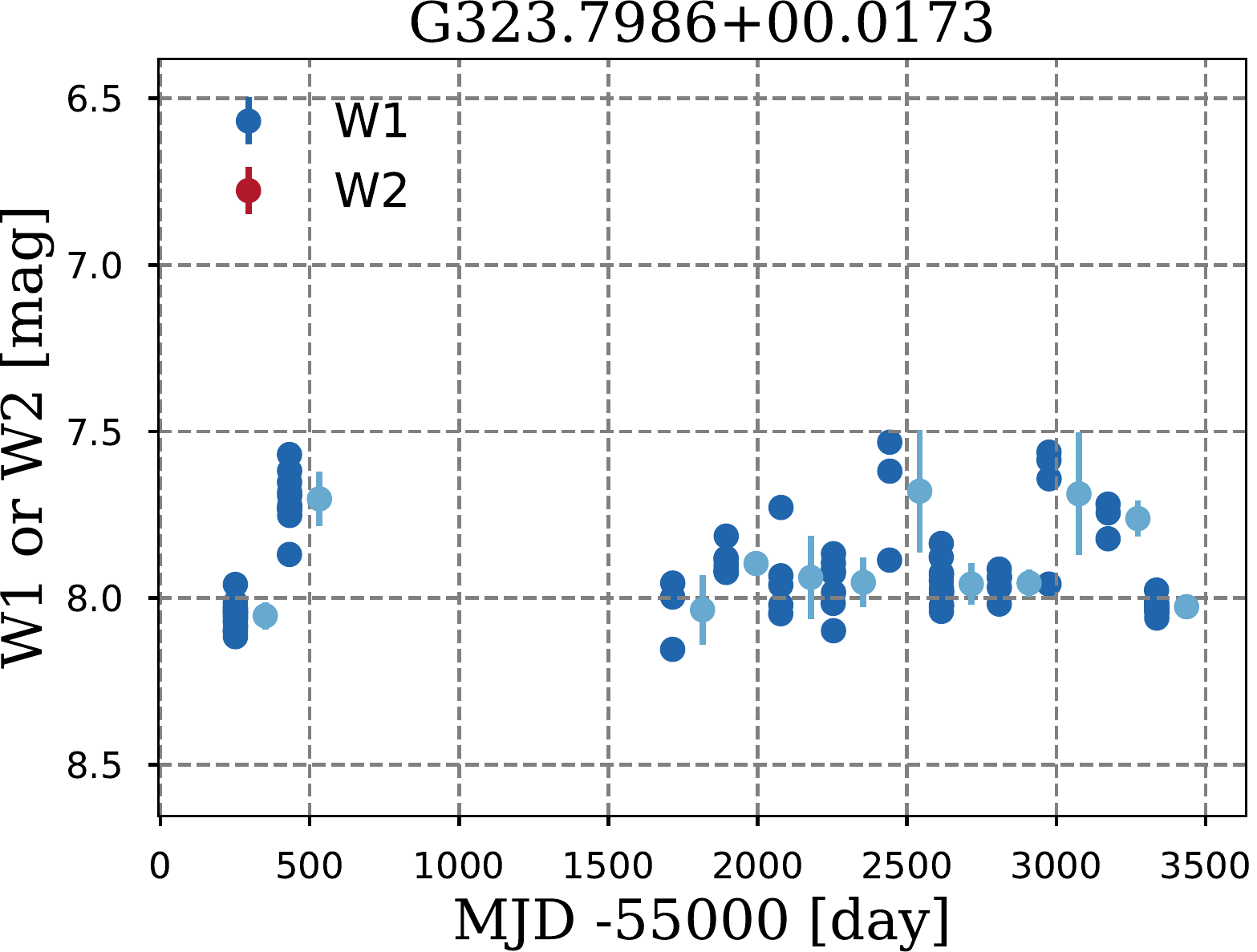}\\
\includegraphics[width=0.33\linewidth]{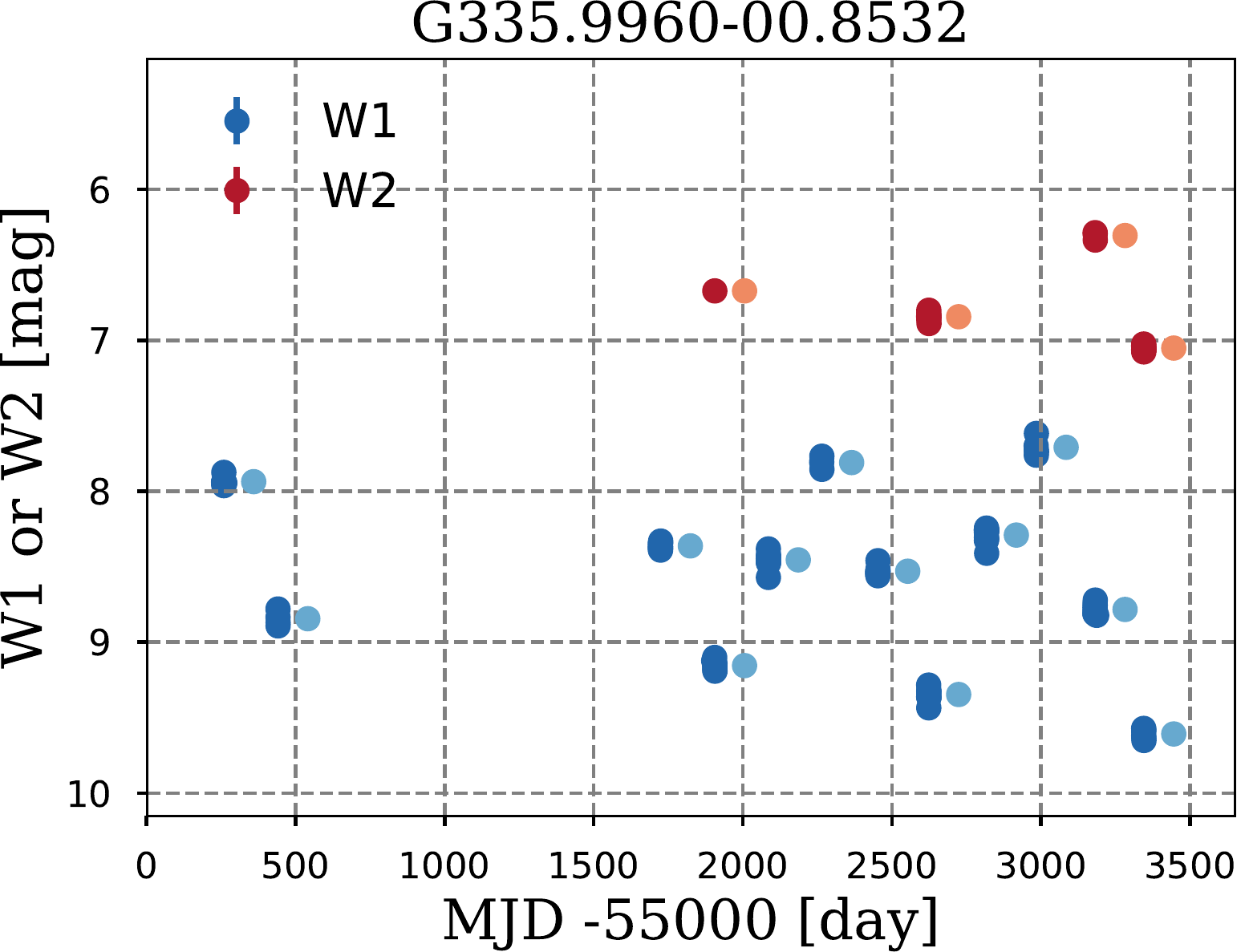}~
\includegraphics[width=0.33\linewidth]{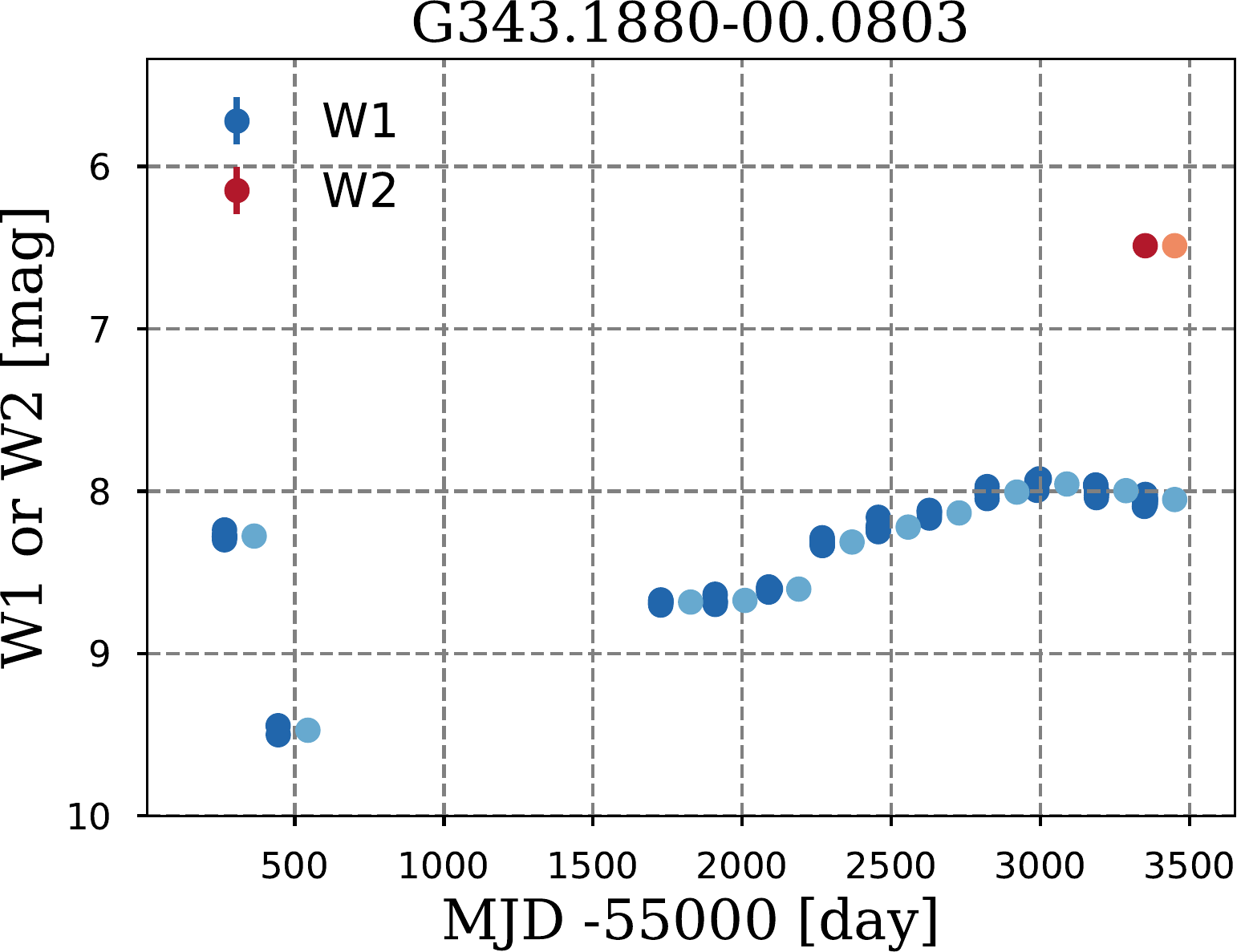}\\
\caption{Infrared light curves of variable sources in W1 and W2 bands. The single exposures with error values are shown in dark-blue (W1) and dark-red (W2), respectively,
and the averaged value in each epoch is shown
in the light-blue (W1) and pink (W2) circles with error bars. The averaged values are shifted to 100 days after the real values for clarity.
}
\label{fig:lc_sec}
\end{center}
\end{figure*}

\textit{WISE} has a 90-min orbit and conducts $\approx 12$ observations of a source over a
 $\approx 1$ day period.
 A given location is observed every 6 months.
 In this study, we refer to this set of $\approx 12$ observations on 1 day as a
 ``1 epoch'' observation.
 The average magnitude and 
 dispersion ($\sigma$) of each epoch observation was derived to determine the significant variability in each MYSO (see also Figure~\ref{fig:lc_sec}).
 As \textit{ALLWISE} covers observations between 2010 January and 2011 February 
 (Modified Julian Date (MJD)$-55000 = 200$--$600$), each source has observations for two epochs.

Using the \textit{ALLWISE} two-epoch observations, 
we searched the variable MYSOs to determine which ones satisfied the four criteria listed below:

\begin{enumerate}

\item Difference in average magnitudes between the two epochs is larger than 
$3\sigma$ in each epoch observation.

\item Difference in average magnitudes between epochs is larger than 0.3 magnitude.

\item $\sigma$ in each epoch is less than 0.3 magnitude.

\item Number of secured photometric data in each epoch is larger than two points.

\end{enumerate}

Out of 331 sources, 12 fulfilled the above criteria.

In addition to applying the criteria of RMS MYSO classification,
we also removed the possible low-mass YSOs.
The RMS project provides the integrated bolometric luminosity of each source using archive MIR/far-infrared (FIR) satellite data and  the associated kinematic distances that they determined \citep{Lumsden13,Urquhart07_co_s,Urquhart08_co_n}.
Objects with luminosity less than 500 $L_{\odot}$ were also removed, as they were likely lower-mass YSOs with corresponding masses of less than about 5 $M_{\odot}$ during the zero-age main sequence \citep{Cox00,Siess00}.
Because MYSOs increase their bolometric luminosity almost monotonically during the mass accretion phase \citep{HO09}, the threshold was set to be slightly below the exact massive-star definition, 8 $M_{\odot}$, so as not to miss potential MYSOs.
This criterion left the final 5 MIR-variable candidates out of 12 sources.

We checked whether there is a source of equal brightness in W1/W2 band within about 15 arcsec or a brighter one within 25 arcsec and confirmed that such source does not exist for all variable candidates in the AllWISE catalog.
We also checked all variable candidates 
by visual inspection using the \textit{WISE} image server\footnote{\url{http://irsa.ipac.caltech.edu/applications/wise/}}, to investigate whether the possible artifacts or diffuse components contributed significantly.
None of the candidates were rejected in these two criteria.
After applying the criteria described above, ultimately, 5 objects remained as MIR-variable MYSOs.

\setlength{\topmargin}{3.0cm}
\setlength{\tabcolsep}{0.020in} 
\floattable
\rotate
\renewcommand{\arraystretch}{1.2}
\begin{deluxetable}{lCCCCCCCCCCccC}
\thispagestyle{empty}
  \tablefontsize{\footnotesize}
  \tabletypesize{\scriptsize}
  \tablecaption{Properties of selected variable sources.  \label{tab:sources}}
  \tablewidth{0pt}
  \tablehead{
\colhead{(1)} & \colhead{(2)} &   \colhead{(3)} &
\colhead{(4)} & \colhead{(5)} &   \colhead{(6)} & \colhead{(7)} & \colhead{(8)} &   \colhead{(9)} &
\colhead{(10)} & \colhead{(11)} & \colhead{(12)} &
\colhead{(13)} & \colhead{(14)}
\\ \hline
\colhead{Object name} & \colhead{RA} & \colhead{Dec}
& \colhead{W1$^{(1)}$} & \colhead{W2$^{(1)}$} & 
\colhead{W3$^{(1)}$} & \colhead{W4$^{(1)}$} 
& \colhead{W1$^{(2)}$} & \colhead{W2$^{(2)}$} &
\colhead{$\Delta$ W1} & \colhead{$\Delta$ W2} & 
\colhead{Variability} &
\colhead{Color} &
\colhead{$L$}
\\
\colhead{} & \colhead{} & \colhead{} & 
\colhead{mag} & 
\colhead{mag} & 
\colhead{mag} & 
\colhead{mag} & 
\colhead{mag} & 
\colhead{mag} & 
\colhead{mag} & 
\colhead{mag} & 
\multicolumn{2}{c}{} &
\colhead{$L_{\odot}$}
  }
  \startdata
\thispagestyle{empty}
G076.1807+00.0619 & 306.012 & 37.610  & 8.57 & 6.85 &  3.65  &  0.67   & - & 6.50 & - & -0.35  & D Pl & BB & $5.45 \times 10^{2}$\\
G269.5205-01.2510 & 136.123 & -48.823  & 8.63 & 6.36 &  2.96  &  -0.84  & 9.29 & - & 0.66 & -   & O & Unclear & $1.0 \times 10^{3}$\\
G323.7986+00.0173 & 232.738 & -56.250 & 8.05 & - & 2.34 & -0.44 & 7.70 & - & -0.35 & - & I Pl & - & $4.58 \times 10^{4}$\\
G335.9960-00.8532 & 248.795 & -48.814 & 7.94 & - &  3.01  &  0.63  & 8.84 & - & 0.91 & -  & P & - & $9.34 \times 10^{2}$\\
G343.1880-00.0803 & 254.642 & -42.832 & 8.28 & - &  3.84  &  1.66   & 9.47 & - & 1.17 & -  & Dip & - & $7.58 \times 10^{2}$\\
    \enddata
    \thispagestyle{empty}
    \tablenotetext{}{
    Notes.
    Columns:
    (1) Object name;
    (2) RA; 
    (3) Dec (J2000);
    (4)--(7) Profile-fitting flux densities
    from W1 to W4 obtained by \textsc{ALLWISE} for the first epoch, while only W1 and W2 are selected data as described in section 2.2 ;
    (8)--(9) Profile-fitting flux densities
    from W1 and W2 obtained by \textsc{ALLWISE} for the
    second epoch using selected data;
    (10)--(11) Infrared (IR) color of the different epochs,
    defined as 
    $\Delta W1 (\Delta W2) = W1^{(1)} (W2^{(1)}) - W1^{(2)} ( W2^{(2)})$.
    (12) Variability type flag as discussed in Section 3.1.1.
    ``P'' represents ``Periodic variability'', ``Dip'' as ``Dipper'', ``I Pl'' as ``Increase with plateau'', ``D Pl'' as ``Decrease with plateau'', and ``O'' as ``Other''.
    (13) Color variability type as discussed in Section 3.1.2.
    ``BB'' represents ``Bluer when Brighter''.
    (14) Estimated bolometric luminosity with 
    units of solar luminosity, obtained from \cite{Lumsden13}.
    }
\end{deluxetable}
\setlength{\topmargin}{0in}

\section{Results and Discussion}\label{sec:results}

Figure~\ref{fig:lc_sec} shows the light curves for the 
five MIR-variable sources of \textit{ALLWISE}, covering observations between 2010 January and 2011 February
(MJD$-55000 = 200$--$600$), and \textit{NEOWISE}, covering observations between 2013 December 13 and 2018 December 13, UTC (MJD$-55000 = 1600$--$3500$).
The variability properties of each variable object are 
summarized in Table \ref{tab:sources}.

\subsection{Classification of Variability}
To investigate the origins of the detected variability, each MYSO is categorized with respect to its time and color flux variation. 

\subsubsection{Time Variation}\label{sec:tv}

Figure~\ref{fig:lc_sec} shows that there are distinct types of
variabilities. To discuss each type of variability, 
the time variation of each MYSO is categorized as ``Periodic'', ``Dipper'',  ``Increase/Decrease with Plateau'', or ``Other''.

In this work, the Lomb$-$Scargle method \citep{Lomb76,Scargle82} was used to search for periodic variability with the AstroPy Lomb$-$Scargle periodogram package \citep{Astropy13}.
This method is widely used to detect periodic variability in unevenly obtained data, which is seen frequently in astronomical observations \citep[e.g.,][]{Morales11,Goedhart14,Sugiyama17}.
Considering that the cadence of the IR observations is roughly
every a half year ($\sim 180$~days), 
the minimum detectable periods should be
larger than the twice of the cadence, i.e., 360~days.
On the other hand, our available dataset spans up to 
nine years
($\sim 3300$~days), indicating that 
the upper-bound of the detectable
period should be smaller than that of the half of the obtained
dataset, i.e., $1650$~days.
Therefore, periods between 360 to 1650 days were searched in \textit{WISE} dataset in this work.
When conducting a Lomb$-$Scargle periodogram analysis, power spectra versus frequency is derived as the calculation result. 
To evaluate the peaks that are significant in the power spectra, 
we used false alarm probability (FAP), which represents the probability of attaining a certain amplitude of power spectra without any periodic component, assuming that the obtained data error consists of only Gaussian noise \citep{VanderPlas18}.
In this work, an FAP of less than $3 \times 10^{-3}$, 3 $\sigma$ detection assuming Gaussian noise distribution, was adopted as the threshold of periodic variability detection.

The criteria for each variability class are given below.

\begin{enumerate}
\item If the FAP of the Lomb$-$Scargle method periodogram is less than $3 \times 10^{-3}$ in either band, the object is categorized as ``Periodic''. 
\item Else if the flux drop between adjacent epoch magnitudes is larger than 1 mag in either data set, the object is categorized as ``Dipper''.
\item Else if the binning fluxes partially increase or decrease while the other epochs are overlapped within $\sigma$, the object is categorized as ``Increase/Decrease with Plateau''.
\item The object which is not categorized as one of the above criteria is labelled as ``Other''.
\end{enumerate}

The classification results are summarized in Table \ref{tab:sources}.
There are two ``Plateau'', one ``Dipper'', one ``Periodic'', and one ``Other'' object.

\begin{figure*}
\begin{center}
\includegraphics[width=0.5\linewidth]{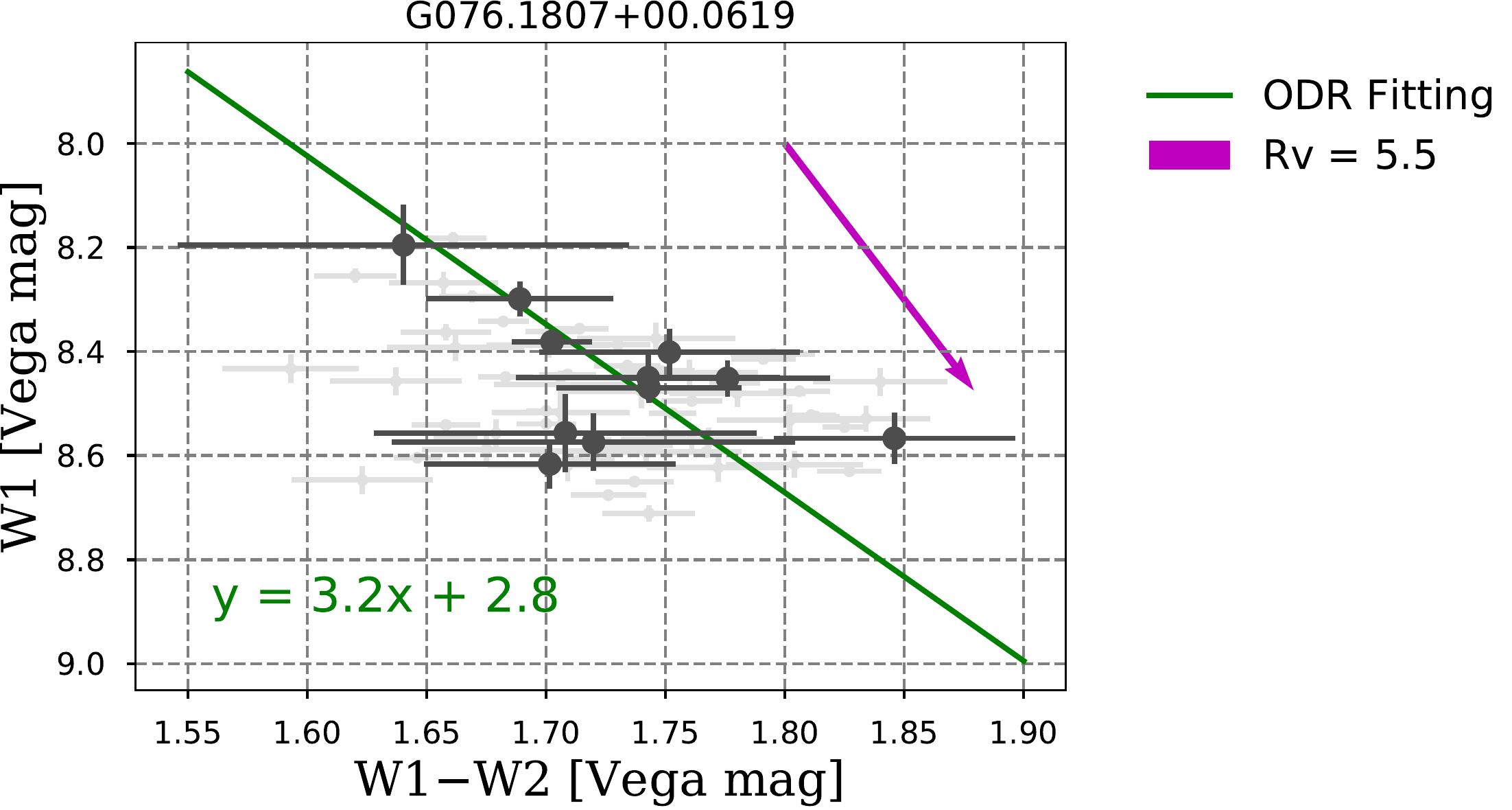}~
\includegraphics[width=0.5\linewidth]{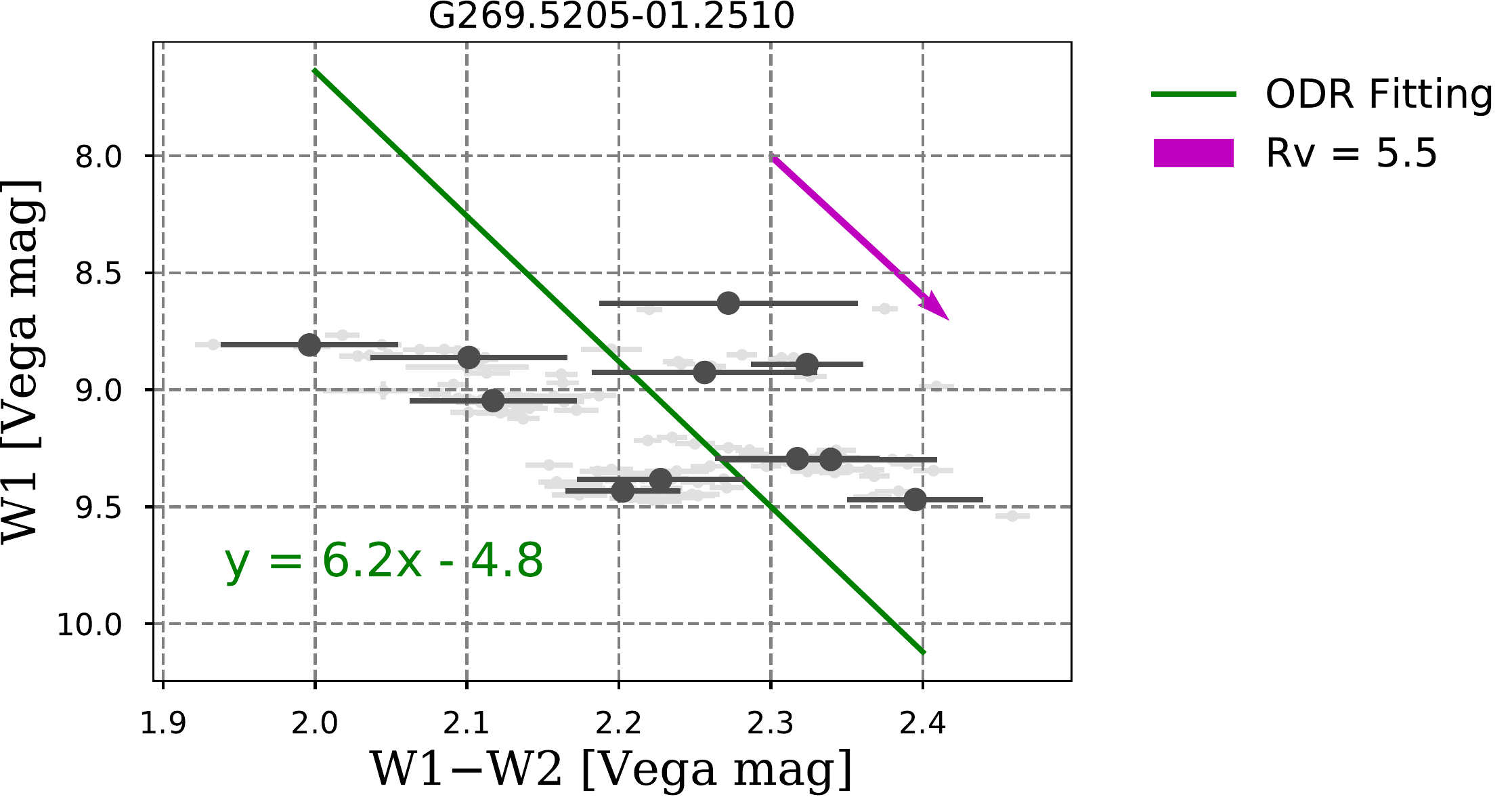}
\caption{
Color magnitude diagram (CMD)
of W1 and W1$-$W2 for variable sources.
The single exposures with their associated error values are
shown in gray, and the average value for each
epoch is shown in black with error bars.
Magenta vector represent extinction vectors at $R_\mathrm{V}=5.5$.
Green line represents ODR fitting result, where $x$ and $y$ are a color of W1$-$W2 and magnitude of W1, respectively.}

\label{fig:CMD}
\end{center}
\end{figure*}

\subsubsection{Color Variation}\label{sec:cv}

A color change during the 
flux variation provides us important information on the possible origins of the variability.
When the flux variation is caused by
either a change in the dust 
temperature of emitting regions, possibly due to mass accretion-rate variation, or 
the level of dust extinction, its color should also vary simultaneously.
In contrast, the occultation by binary stars or planets in the line-of-sight affects the normalization of the flux,
but does not produce a color change between W1 and W2.

Figure~\ref{fig:CMD} shows the two
CMDs
of W1$-$W2 and W1
for our MIR variability sources which have selected W1 and W2 data more than four at the same observation time, G076.1807+00.0619 and G269.5205-01.2510.
According to the MIR color variability analysis in YSOVAR project \citep{Gunther14}, we used Orthogonal Distance Regression (ODR) method to derive a slope of each CMD.
When analyzed data of W1$-$W2 and W1 is noise limited, then artificial slope of 45 degree is obtained \citep{Gunther14}.
Therefore, we derive slopes of CMD to investigate whether obtained values could be 45 degree within ODR fitting error.

Our fitted slopes of G076.1807+00.0619 and G269.5205-01.2510, are over-plotted in Figure~\ref{fig:CMD} and their equations are  $y =(3.2 \pm 1.3)x+(2.8 \pm 2.3)$ and $y = (6.2 \pm 5.0)x - (4.8 \pm 11)$ respectively, where $x$ and $y$ are a color of W1$-$W2 and a magnitude of W1.
In Figure~\ref{fig:CMD}, an extinction vector at $R_\mathrm{V}=5.5$, typical value at both interstellar medium and dense molecular clouds in 3-8 $\micron$ \citep{WD01,WLJ14}, is also shown.
G076.1807+00.0619 object shows a bluer (redder) color when the flux is increased (decreased) and 
the obtained slope is steeper than the 45 degree including fitting error.
Therefore, the obtained slope shows ``Bluer when Brighter (BB)'' color trend.

On the other hand, 
the slope of G269.5205-01.2510 has large error in the fitting. Therefore, we could not find a statistically significant color trend for G269.5205-01.2510.

\subsection{Fraction of MIR-variable MYSOs and Comparison with Lower-mass YSOs}

Our results show that the fraction of 
variable MYSOs is 5/331, or equal to 1.5\%,
which seems to be much smaller than that found by lower-mass variables such as 70 \% at YSOVAR \citep{Morales11}.
However, year-scale variables are rare even in lower-mass YSOs, because most of variability in lower-mass YSOs are related to rotation and hence its periods are a few days or less \citep{Rebull14,Morales11}.
The fraction of year-scale variability ($\Delta$ [3.6] $\gtrsim$ 0.05 mag) in YSOVAR reduces into roughly 20-30 \% \citep{Rebull14}.
While the value is still lager than our result of 1.5 \% ($\Delta\left( [3.4]/[4.6] \right) > 0.3$ mag),
our detection-threshold requires 
higher amplitude with $>0.3$~mag, 
which might miss possible variabilities with smaller
amplitude of $0.05 < \Delta\left( [3.4]/[4.6] \right) < 0.3$ mag as seen in the sample of YSOVAR.
Actually, amplitudes of year-scale flux variations are generally weaker than those of day-scale ones in lower-mass YSOs according to The Research Of Traces Of Rotation (ROTOR)-program in V band observations \citep{Grankin07} and if this is the case with MYSOs, current selection method may miss many year-scale variables, although such ambiguous variables are out-of-scope in this paper.
Thus, the obtained variability fraction
of 1.5\% in this study could be considered as an lower-limit.

\subsection{Characteristics of Variability in MYSOs}
The cross matrix of 
time variability (see Section~\ref{sec:tv}) 
and color variability (see Section~\ref{sec:cv}) 
is shown in Table \ref{tab:trend}.
In this table, while about three objects were not able to conduct ODR analysis and one object does not show clear color trend in ODR analysis, 
one ``Plateau'' object shows ``BB''
color tendency, suggesting that an
origin of those MIR variabilities of either
the extinction and/or temperature change 
of the MIR emission region of the source, such as change in accretion rate.

G335.9960-00.8532 showed clear periodic variability in the W1 band, with an FAP of less than $3 \times 10^{-3}$; however, no color variability was available because of the saturation of W2 band data.
Detailed stellar parameters and possible origins of periodic variability are discussed in Section \ref{sec:periodic}.

\begin{deluxetable}{cccccc}
\tablecaption{Classification based on the time and color variations of MYSOs in this study.\label{tab:trend}}
\tabletypesize{\footnotesize}
\tablewidth{0pt}
\tablehead{
\colhead{}&
\colhead{BB}&
\colhead{Unclear}&
\colhead{-}&
\colhead{Total}
}
\startdata
Periodic & 0 & 0& 1 & 1 \\  
Dipper & 0 & 0 & 1 & 1 \\
Plateau & 1 & 0 & 1 & 2 \\
Other & 0 & 1 & 0 & 1 \\
Total & 1 & 1 & 3 & 5 \\
\enddata
\tablecomments{
``BB'' represents `Bluer when Brighter' color variability ; ``-'' represents unavailable CMD data.
Detailed classification criteria are described in Sections \ref{sec:tv} and \ref{sec:cv}. 
}
\end{deluxetable}

\subsection{Possible Origins of Flux and MIR Color Variation}

\subsubsection{Plateau class with BB color}

Long-term variability has been studied in lower-mass YSOs.
ROTOR-program shows year-scale variability in the optical \textit{UBVR} bands 
for classical T Tauri stars (CTTs) from their 20-year monitoring program \citep{Grankin07}.
Approximately 75\% of the long-term variability 
observed in those CTTs shows moderate and almost monotonic flux variation with $\Delta V \leq 1.6$ mag, and color variations are consistent with the interstellar extinction vector.
This suggests that the main cause of long-term variability in CTTs is
from a difference in line-of-sight extinction or variation in the mass accretion rate \citep{Grankin07}.  
The color variability trends which are consistent with the interstellar extinction vector or somewhat much more colorless ones are also generally found in MIR variability study of lower-mass YSOs \citep{Gunther14}. 
They also suggested the difference in line-of-sight extinction or variation in the mass accretion rate.
Although 90\% quantile of obtained $A_\mathrm{V}$ corresponds to  $\Delta V \leq $ 10 mag and this value is larger than that in the ROTOR-program, this may be the result that much younger objects included in the work of \citet{Gunther14}.

A similar trend is 
evident in G076.1807+00.0619, which shows ``Plateau'' time-variability MYSO associated with the ``BB'' color variability; this suggests that these long-term variabilities may be caused by the transition of the line-of-sight extinction, such as the rotation of the accretion disk with a non-axisymmetric angular density distribution 
or change in mass accretion rate possibly caused by binary interaction \citep{Araya10} due to high close binary rate in massive stars \citep{Chini12}, or gravitational instability of accretion disk due to high accretion rate \citep{Matsushita17,Meyer19}.
Assuming that the accretion disk of those MYSOs
is Keplerian rotation with a central MYSO mass of $\sim10 M_{\odot}$ and bolometric luminosity of $\sim5000 L_{\odot}$, 
the dynamical timescale of the observed variation of $\sim 1000$~days in this study corresponds to a radius of $R_{\rm dyn}\sim10$~AU, which is five-fold larger than the dust sublimation radius $R_{\rm sub}$ of $\sim 2$~AU ($R_{\rm dyn} \sim 5 R_{\rm sub}$).
Further spectroscopic monitoring of hydrogen recombination lines related to the mass accretion rate \citep[e.g.,][]{Alcala14} is critical to
resolve the degeneracy of the two possible origins between the extinction and the change in accretion rate.

\begin{figure*}
\begin{center}
\includegraphics[width=0.75\linewidth]{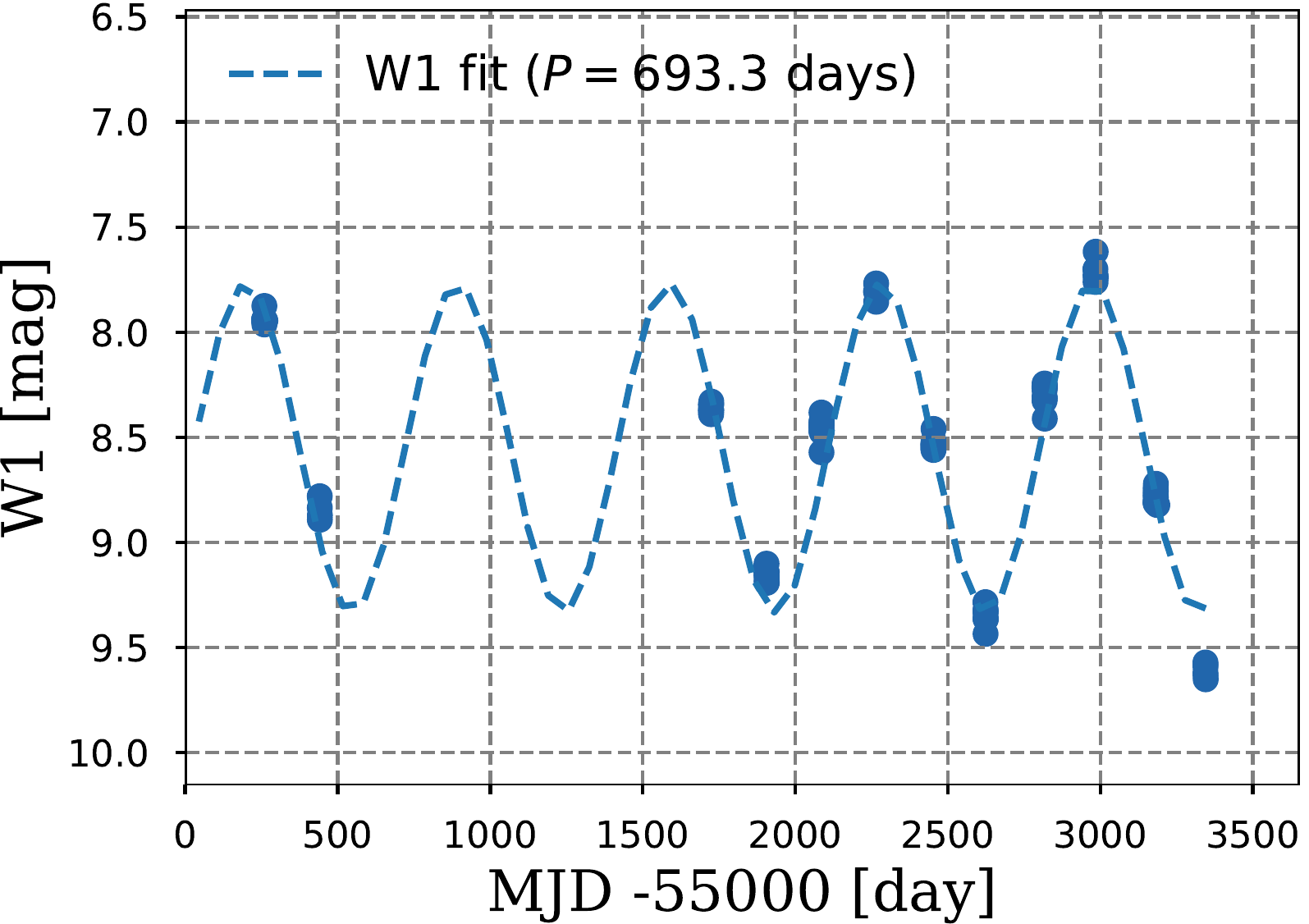}~
\caption{Best-fitted period of G335.9960-00.8532 in W1 band. Periods of 693.3 days were derived.}
\label{fig:periodic}
\end{center}
\end{figure*}

\subsubsection{Dipper Class}

One detected ``Dipper'' class object, G343.1880-00.0803, shows a sudden 
flux decrease with subsequent flux recovery.
The above characteristics well resemble the 
``dipper'' objects in lower-mass YSOs \citep{Gunther14,Cody14,Bibo90,Waters98},
exhibiting a flux decrement due to the sharp line-of-sight extinction, while CMD of G343.1880-00.0803 was unavailable.

\subsubsection{Other class}

The object, G269.5205-01.2510, is categorized as ``Other'' in terms of time variability.
This would be mainly associated with the insufficient cadence or the duration of observations.
While the origin of the \textit{WISE} MIR color trend in this object is not clear yet, the K$-$W1 color referred from RMS survey is reported to be very large, about 4-5 magnitudes \footnote{\url{http://rms.leeds.ac.uk/cgi-bin/public/RMS_SUMMARY_PAGE.cgi}}. Again, its origin is not clear at the current stage.
Thus, further observations with higher cadence and/or longer duration in the NIR and MIR should reveal a hidden variability trend with a shorter timescale.

\begin{figure}
\begin{center}
\includegraphics[width=1.0\linewidth]{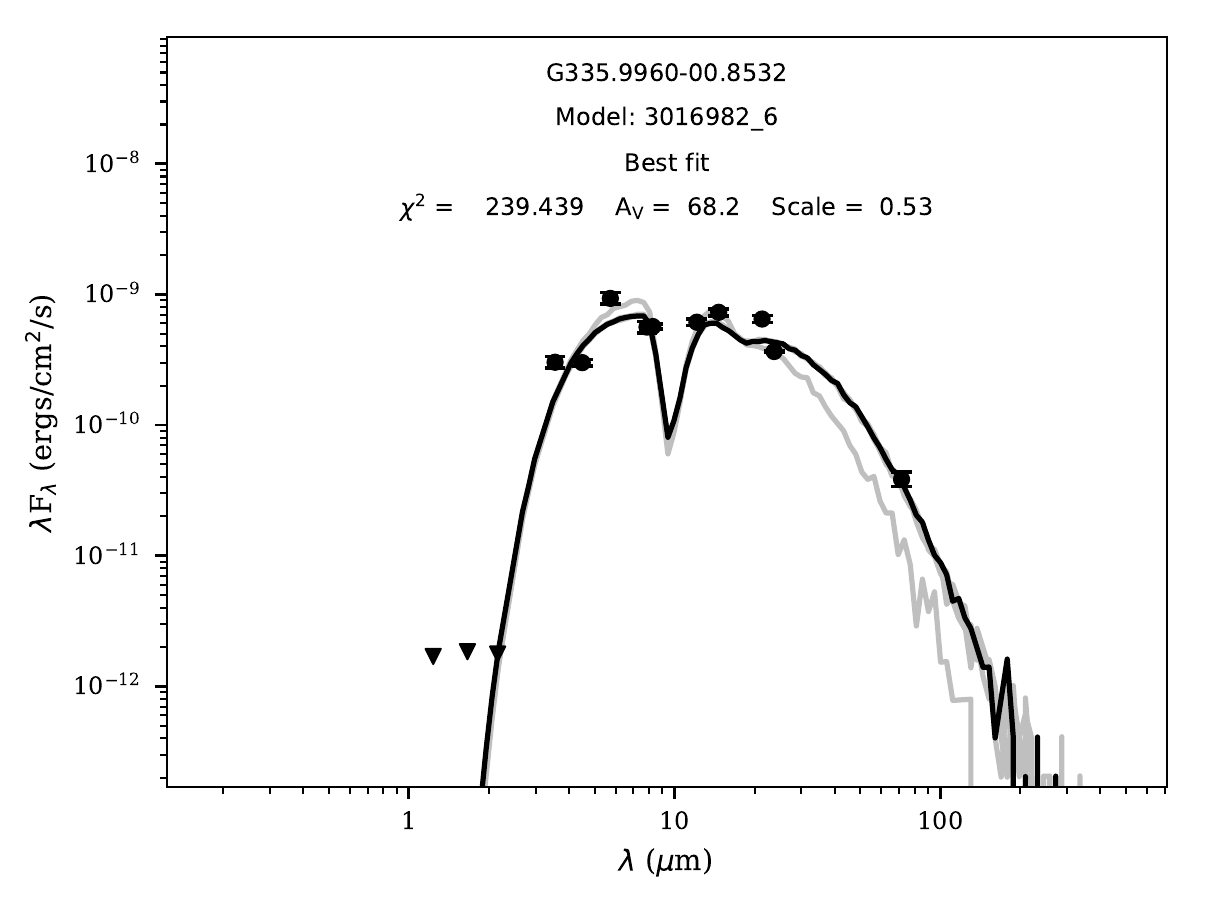}~
\caption{Obtained spectral energy distribution (SED) for G335.9960-00.8532 and 
the best-fitting result. The black points with the error bars 
correspond to the obtained infrared (IR) photometries from 2MASS, MSX,
\textit{Spitzer}, and \textit{Herschel}/PACS.
As all 2MASS bands are non-detections, we show them
as upper bounds, with the triangles of the J,  H, and K bands.
The best-fitting SED is indicated by a solid black line.
}
\label{fig:SEDfit}
\end{center}
\end{figure}

\subsubsection{Periodic variable MYSO candidate G335.9960-00.8532}\label{sec:periodic}

According to the periodic analysis described in section~\ref{sec:tv}, a clear periodic variability is detected in G335.9960-00.8532
(hereafter, G335).
G335 has a period of 693.3 days in the W1 band, with an FAP of $5.6 \times 10^{-4}$.
Assuming its light curve is approximated as sinusoidal curve,
the best-fitting modeled light-curve is overlaid in Figure \ref{fig:periodic}.
While the periodicity is clearly shown in Figure~\ref{fig:periodic}, the CMD and hence the MIR color trend was not available for G335 due to saturation in W2 band.

In further examination of the  possible origins of periodic variability, we estimated stellar and accretion disk parameters, using a spectral energy distribution (SED) fitting based on the MYSO models, including various combinations of stellar parameters, such as extinction, total bolometric luminosity, stellar mass, effective temperature, and disk mass \citep{Robitaille07}.

The SED of G335 consists of the archival data of MSX (RMS project), \textit{Spitzer}, and \textit{Herschel} (Hi-GAL program) \citep{Lumsden13,wri10,Molinari16}
by cross-matching the coordinates obtained by the RMS catalog
with the searching radius of 5~arcsec.
SED fitting results are shown in Figure~\ref{fig:SEDfit}, and 
the derived stellar parameters are summarized in Table~\ref{tab:G335para}.
Because the distance to the object
is obtained from the kinematic near-side distance
\citep{Urquhart07_co_s}, the obtained
bolometric luminosity and the other 
luminosity-related parameters correspond to the
lower bound.
The associated stellar parameters indicate
that G335 contains a central MYSO of 10.5 $M_{\odot}$, with an associated small mass disk of $\sim 4.2 \times 10^{-3} M_{\odot}$.
This indicates that the evaporation is due to the irradiation of the central MYSO or late-phase MYSO evolution.
Although the model fitting shows that
the temperature of the central MYSO is well above $10^5$ T, the \ion{H}{2} region cannot be detected based on high-spatial resolution radio observation \citep{Urquhart07_hii_s}.
One possible origin is that during the late-phase MYSO evolution, as the central engine begins to blow away the surrounding dust region, the central \ion{H}{2} region remains compact and sufficiently dense, such that it cannot be detected by radio observation.

Some origins may provoke periodic MIR flux variation, as observed in G335. 
The stellar rotation with cool/hot spots is a well-observed phenomenon in lower-mass YSOs, as periodic day-scale variability \citep{Morales11}.
However, because the observed period for G335 is 
much longer at $>100$~days, an origin of stellar rotation is unlikely \citep{Rosen12}.
The other interesting idea is a close binary origin. Because main sequence massive stars frequently have binary stars, 
a close binary system could be easily realized for G335  \citep{Chini12}.
However, it is difficult for occultation of the binary system to achieve a sinusoidal-like light curve, as observed in Figure~\ref{fig:periodic}.

Close interaction of a MYSO binary system could produce the periodic change in mass accretion rate and thus a resulting sinusoidal-like light curve as suggested in \cite{Araya10}. 
Since the change in mass accretion rate causes ``BB'' color variability in the MIR \citep{Gunther14}, multi-color monitoring observations are necessary to evaluate this hypothesis.
Similarly, recent theoretical studies indicate that change in accretion rate caused by either a disk-fragmented companion in disk or gravitational disk instability can occasionally make sinusoidal-like light curves \citep{Matsushita17,Meyer19}, although the current spatial resolution of 
those simulations is not high enough to resolve an inner 10-AU region. Therefore, our results would motivate such simulation studies to explore more finer spatial resolutions by resolving the year-scale flux variation.

Another possible scenario is the 
periodic/aperiodic change in extinction in the line of sight, such as 
the non-axisymmetric dust density distribution in the rotating accretion disk.
The CMD diagram gives us to judge whether
it is also the case for G335, while the CMD is not available because of the saturation of W2 band.

Recent theories have proposed that
stellar pulsation occurs with late-phase accreting MYSOs \citep[e.g.,][]{Inayoshi13};
the obtained SED-fitting result for G335 is consistent with the scenario proposed above.
In the prediction, a $\sim 100$-day-scale periodic variation that follows a period-luminosity relation was assumed; thus, the observed period
for G335 should be the luminosity of 
$2.8 \times 10^{5} L_{\odot}$, 
which is one order of magnitude 
larger than the bolometric luminosity
of $7.6 \times 10^3 L_{\odot}$ obtained by the SED fitting.
When the far-side distance, 12.1 kpc, was employed, the bolometric luminosity became $1.0 \times 10^{5} L_{\odot}$, consistent with the period-luminosity relation within a factor of three.
However, it would be difficult to observe the MYSO located at 12.1 kpc beyond the Galactic center region, even in the MIR and FIR regions.
Because the current model of
\cite{Inayoshi13} assumes a
constant accretion rate, it 
may affect the predicted relation.
From a more dynamic perspective, a time-dependent accretion model for stellar pulsation may be necessary to fill in the gap between theory and observation for G335.

Follow-up observations in the NIR and MIR regimes of G335 are planned for upcoming ground telescope observations through The University of Tokyo Atacama Observatory (TAO)/Simultaneous-color Wide-field Infrared Multi-object Spectrograph (SWIMS) and Mid-Infrared Multi-mode Imager for gaZing at the UnKnown Universe (MIMIZUKU) \citep{Motohara16,Kamizuka16}.
To evaluate variations in the mass accretion rate and dust extinction during flux variability, 
the NIR spectroscopic monitoring of hydrogen recombination lines and the MIR narrow-band imaging of dust silicate features are critical.
Monitoring observations in the 10-$\micron$ band is currently difficult, due to a lack of access to the MIR instrument.
The new ground telescope, TAO, will enable the acquisition of $>1$~year-scale monitoring data.

\section{Summary and Conclusions}
We conducted a systematic search
for long-term ($>1$~yr scale) 
MIR ($\lambda>3 \mu$m) variability
of MYSOs utilizing all-sky 
 \textit{ALLWISE} and \textit{NEOWISE} 
data.
This provided a periodic MIR
light curve at the bands W1 ($3.4 \mu$m)
and W2 ($4.6 \mu$m), 
with a cadence of every 6 month
during the 9-year-long coverage period.
We cross-matched between
the \textit{WISE} catalogs and 806 MYSOs obtained by the RMS survey, the largest MYSO catalog in the MIR band with wide wavelength coverage, including those of the radio band.
After the cross-matching process
and the variability selection,
we found five MIR-variable sources.

We categorized the time variability classes based on the obtained 9-year-long MIR light curves
into periodic, dipper, plateau, and others.  
G335.9960-00.8532 (G335) shows 693.3-day periodic variability in W1 band, two sources show a plateau-like light curve, and one shows ``eruption''-like, and one is categorized as ``other''.

For the ``plateau'' class, one out of the two objects showed ``BB'' MIR color trends, suggesting that the MIR variability originates from a change in the accretion rate and/or dust extinction, which is also the case with many lower-mass MIR-variable YSOs. 
The ``dipper'' object G343.1880-00.0803 showed a drastic flux change, similar to ``dipper'' objects in lower-mass YSOs, whereas its CMD is not available and possible origin is not currently clear.

For G335, the only ``periodic'' class,
we applied SED model fitting to estimate the stellar and accretion disk parameters.
Our results indicated that G335 is in the phase of a relatively evolved MYSO; thus, we may be witnessing the very early stages of a hyper- or ultra-compact \ion{H}{2} region.

The remaining one objects G269.5205-01.2510, categorized as ``other'' class and showing no clear color trend because of the large scatter, are good candidates for future follow-up observations with various cadence and multi-band observations, including spectroscopy to determine the time and color MIR variabilities to resolve their origins.

\acknowledgments

We thank Kohei Inayoshi and Kei Tanaka for discussions and feedback, which improved our paper.
We appreciate the support of Sizuka Nakajima for collection of important data.
This study benefited from financial support from the Japan Society for the Promotion of Science (JSPS) KAKENHI program (18K13584; KI),
and the Japan Science and Technology Agency (JST) grant ``Building of Consortia for the Development of Human Resources in Science and Technology'' (KI).
This publication makes use of data products from the Near-Earth Object Wide-field Infrared Survey Explorer (NEOWISE), which is a project of the Jet Propulsion Laboratory (JPL)/California Institute of Technology. NEOWISE is funded by the National Aeronautics and Space Administration (NASA).
This research made use of data products from the Midcourse Space Experiment. Processing of the data was funded by the Ballistic Missile Defense Organization with additional support from NASA Office of Space Science. This research has also made use of the NASA/ IPAC Infrared Science Archive, which is operated by the Jet Propulsion Laboratory, California Institute of Technology, under contract with the National Aeronautics and Space Administration.
Finally, this paper made use of information from the Red MSX Source survey database at http:\slash \slash rms.leeds.ac.uk\slash cgi-bin\slash public\slash RMS\_DATABASE.cgi which was constructed with support from the Science and Technology Facilities Council of the UK.

\facilities{
2MASS, \textit{WISE}, \textit{Spitzer}, \textit{Herschel}, \textit{MSX} }

\software{astropy \citep{Astropy13}, Matplotlib \citep{hun07}, Pandas \citep{mck10}}

\begin{deluxetable}{lcc}
\tablecaption{
The Main Parameters for the Best-fitting Models of G335.9960-00.8532\label{tab:G335para}}
\tabletypesize{\small}
\tablecolumns{2}
\tablenum{3}
\tablewidth{5pt}
\tablehead{
\colhead{Col.} &
\colhead{Parameter} &
\colhead{G335}
}
\startdata
\multicolumn{3}{c}{General}\\
\hline
(1) & $A_{\rm V}$ & $69.7$\\
(2) & Evol. Stage & Disk\\
(3) & $L_{\rm tot}/L_{\odot}$ & $7.6\times10^{3}$\\
\hline
\multicolumn{3}{c}{Central source}\\
\hline
(4) & $M_{\star}/M_{\odot}$ & $10.6$ \\
(5) & $R_{\star}/R_{\odot}$ & $4.13$ \\
(6) & $T_{\star}/{\rm K}$ & $2.65\times10^{4}$\\
\hline
\multicolumn{3}{c}{Disk component}\\
\hline
(7) & $M_{\rm disk}/M_{\odot}$ & $4.2\times10^{-3}$ \\
(8) & $R_{\rm disk,in}/{\rm AU}$ & $61.4$ \\
(9) & $R_{\rm disk, out}/{\rm AU}$ & $6.39\times10^3$\\
\enddata
\tablecomments{Columns: (1) Foreground dust extinction.
(2) Evolutional stage of the YSOs.
‘‘Disk’’ refers to sources that are surrounded only by a circumstellar disk.
(3) Total bolometric luminosity of G~335.
(4) Central stellar mass.
(5) Central stellar radius.
(6) Central stellar temperature.
(7) Disk mass.
(8) Disk inner radius.
(9) Disk outer radius.
}
\end{deluxetable}

\bibliographystyle{apj}
\bibliography{apj-jour,ref.bib}

\end{document}